# Spike-Timing Dependent Plasticity Effect on the Temporal Patterning of Neural Synchronization

**Joel Zirkle[1], Leonid L Rubchinsky[1,2]***

[1] Department of Mathematical Sciences, Indiana University Purdue University Indianapolis, Indianapolis, IN, USA

[2] Stark Neurosciences Research Institute, Indiana University School of Medicine, Indianapolis, IN, USA

**\* Correspondence:**
Leonid Rubchinsky
lrubchin@iupui.edu

## Abstract

Neural synchrony in the brain at rest is usually variable and intermittent, thus intervals of predominantly synchronized activity are interrupted by intervals of desynchronized activity. Prior studies suggested that this temporal structure of the weakly synchronous activity might be functionally significant: many short desynchronizations may be functionally different from few long desynchronizations even if the average synchrony level is the same. In this study, we used computational neuroscience methods to investigate the effects of spike-timing dependent plasticity (STDP) on the temporal patterns of synchronization in a simple model. We employed a small network of conductance-based model neurons that were connected via excitatory plastic synapses. The dynamics of this network was subjected to the time-series analysis methods used in prior experimental studies. We found that STDP could alter the synchronized dynamics in the network in several ways, depending on the time scale that plasticity acts on. However, in general, the action of STDP in the simple network considered here is to promote dynamics with short desynchronizations (i.e. dynamics reminiscent of that observed in experimental studies). Complex interplay of the cellular and synaptic dynamics may lead to the activity-dependent adjustment of synaptic strength in such a way as to facilitate experimentally observed short desynchronizations in the intermittently synchronized neural activity.

## 1    Introduction

Synchronization of neural activity in the brain is involved in multiple neural functions (e.g. Buzsáki and Draguhn, 2004; Fell and Axmacher, 2011; Fries, 2015; Harris and Gordon, 2015). Neural synchronization that is either too strong or too weak may be one of the neurophysiological factors behind symptoms of several disorders such as Parkinson's disease and schizophrenia (Schnitzler and Gross, 2005; Uhlaas and Singer, 2006; Oswal et al., 2013; Pittman-Polletta et al., 2015). Thus, the synchronization of neural activity is a ubiquitous phenomenon. In the rest state, the strength of this synchronization is usually moderate. This means that the intervals of stronger synchrony are interspersed with desynchronized intervals. This is probably not surprising given the plausibility of the very general nature of the transient character of neural activity (Rabinovich et al., 2008).

Recent developments in time-series analysis allowed for the exploration of the temporal patterning of synchronized activity in brain dynamics on very short time-scales. Studies of different brain signals in different conditions and species suggest an apparently universal feature: synchronous activity is interrupted by very short (although potentially numerous) intervals of desynchronized dynamics (as opposed to few longer desynchronized episodes). This phenomenon was observed in the synchrony between local field potentials (LFPs) and spikes in different parts of the basal ganglia and EEG in Parkinson's disease (Park et al., 2010; Ratnadurai-Giridharan et al., 2016; Ahn et al., 2018), in synchronization between LFPs recorded in the prefrontal cortex and hippocampus of normal and amphetamine-sensitized mice (Ahn et al., 2014), in EEG of healthy human subjects (Ahn and Rubchinsky, 2013), and in EEG in autism spectrum disorders (Malaria et al., 2020). The differences in the temporal patterning are correlated with certain behavioral features but the prevalence of short desynchronizations persisted nevertheless (Ahn et al., 2014, 2018; Malaia et al., 2020). Therefore, short desynchronizations may be functionally important and the properties and mechanisms of desynchronization durations merit exploration.

These observations of the persistence of short desynchronizations naturally suggests the question about the biological mechanisms behind this phenomenon. The modeling study (Ahn and Rubchinsky, 2017) suggested one possible mechanism: the short desynchronization dynamics was promoted by the substantial difference in the timescales of spike-producing sodium and potassium currents. The relative slowness of the potassium delayed-rectifier current may be one of the reasons for why short desynchronizations are observed in different neural systems. However, there may also be other mechanisms. This paper is aimed at the exploration of one potential mechanism related to synaptic plasticity. We use computational modeling to explore how spike-timing dependent plasticity (STDP) can affect the temporal patterning of neural synchrony on short timescales.

STDP is a very common neural phenomenon with potentially multiple effects on neural synchronization. In particular, a synapse whose conductance is modulated by STDP can enhance neural synchrony (Nowotny et al., 2003; Cassenaer and Laurent, 2007; Ratnadurai-Giridharan et al., 2015). We use a simple neural network of two conductance-based model neurons coupled via excitatory synapses with STDP and apply the same time-series analysis techniques as were used in the prior experimental studies. While this model network can hardly adequately model field potentials recorded in some of the experimental studies mentioned above, it serves as a simple model system exhibiting rich synchronization dynamics, which is substantially modulated by synaptic plasticity. Numerical analysis of this model shows that STDP may affect not only the strength of synchronization, but also the temporal patterning of synchronization, with an ability to facilitate the short desynchronizations dynamics observed in experiments.

## 2    Methods

### 2.1    Neuronal and synaptic modeling

We utilize the network model from (Ahn and Rubchinsky, 2017) except that the synapses are plastic in this study. The model is described below.

The neurons are modeled using a two-dimensional conductance-based model of a Hodgkin-Huxley type that is mathematically equivalent to the Morris-Lecar model (Izhikevich, 2007; Ermentrout and Terman, 2010). The sodium conductance is assumed to activate instantaneously and to have no inactivation, while the potassium conductance is controlled by its gating variable and so varies dynamically.

$$\frac{dv}{dt} = -I_{Na} - I_K - I_L - I_{syn} + I_{app}$$

$$\frac{dw}{dt} = \frac{w_\infty(v) - w}{\tau(v)}$$

Here $v$ is the neuron's transmembrane potential and $w$ is the gating variable for the potassium current. The synaptic current between neurons, $I_{syn}$, is given below and $I_{app}$ is a constant input current to each neuron to control the frequency of spiking. The sodium, potassium, and leak currents are:

$$I_{Na} = g_{Na} m_\infty(v)(v - v_{Na})$$

$$I_K = g_K w(v - v_K)$$

$$I_L = g_L(v - v_L)$$

$g_{Na}$, $g_K$, and $g_L$ are the maximal conductances for the sodium, potassium and leak currents, respectively. The steady-state values for the gating variables of the sodium and potassium currents are:

$$m_\infty(v) = \frac{1}{1 + exp\left(-2\frac{v - v_{m1}}{v_{m2}}\right)}$$

$$w_\infty(v) = \frac{1}{1 + exp\left(-2\frac{v - v_{w1}}{\beta}\right)}$$

The voltage-dependent activation time constant of the potassium current is:

$$\tau(v) = \frac{1}{\epsilon} * \frac{2}{exp\left(\frac{v - v_{w1}}{2\beta}\right) + exp\left(\frac{v_{w1} - v}{2\beta}\right)}$$

All synapses are excitatory, and the synaptic current to neuron $i$ is given by:

$$I_{syn,i} = g_{syn}(v_i - v_{syn})\sum_{j \neq i} s_j$$

Where $g_{syn}$ is the maximal conductance of the synapse (i.e. the synaptic strength), and $s_j$ is the synaptic variable for neuron $j$ and the summation is taken over all neurons that are connected to the $i$-th neuron. The synaptic variable $s$ is governed by:

$$\frac{ds}{dt} = \alpha_s(1 - s)H_\infty(v - \theta_v) - \beta_s s$$

$H_\infty$ is a sigmoidal function whose input is the presynaptic neuronal voltage:

$$H_\infty(v) = \frac{1}{1 + exp\left(-\frac{v}{\sigma_s}\right)}$$

The values of cellular and synaptic parameters are the same as used in (Ahn and Rubchinsky, 2017): $g_{Na} = 1$, $g_K = 3.1$, $g_L = 0.5$, $v_{Na} = 1$, $v_K = -0.7$, $v_L = -0.4$, $v_{m1} = -0.01$, $v_{m2} = 0.15$, $v_{w1} = 0.08$, $\beta = 0.145$, $I_{app} = 0.045$, $\varepsilon_1 = 0.02$, $\varepsilon_2 = 1.2\varepsilon_1$, $v_{syn} = 0.5$, $\propto_s = 5$, $\beta_s = 0.2$, $\theta_v = 0.0$, $\sigma_s = 0.2$.

STDP modeling follows (Zhigulin et al., 2003). If neuron $i$ spikes at time $t_i$ and neuron $j$ spikes at time $t_j$, then the strength of the synapse from neuron $i$ to neuron $j$ is additively updated by the amount

$$\Delta g_{syn} = sgn(\Delta t)Aexp(-k|\Delta t|)$$

where $\Delta t = t_j - t_i$. The synaptic conductance from neuron $j$ to neuron $i$ is simultaneously updated by an equal, but opposite, amount. While the additive update rule does not necessary need to be symmetric (as it is here), there is experimental evidence supporting the nature of the update, see for example (Zhang et al., 1998, Feldman, 2012). We varied the values of our plastic parameters, in particular $A \in [0.0001,0.01]$, $k \in [0.01,50]$. The synaptic conductance is bounded below by zero.

## 2.2 Numerical Implementation

The system of differential equations was solved numerically in Python using the built-in odeint function from the SciPy module (v.1.4.1). This function implements either the Adams method or a backward differentiation formula (BDF) method depending on the stiffness of the problem. The solution was reported at multiples of the time step $dt = 0.1$ (assuming the time units are milliseconds), however the function uses an adaptive step size and there was no lower bound on the length of the intermediate time steps that may be used (similarly, there was no upper bound restriction on the number of intermediate steps that were taken). The absolute and relative tolerances for the method were kept at the default value of $1.49 \times 10^{-8}$. While the solution depends on the initial conditions, its statistical properties (such as the firing rate, synchrony pattern characteristics etc.) do not. The system was solved on the time interval [0,25000], the first 20% of the time-series was removed from analysis. To implement plasticity, the integration was paused after each time step and, if necessary, the synaptic strength was updated. Specifically, the voltage threshold to define an action potential was set at 0.2.

## 2.3 Synchronization analysis

The time-series analysis of synchronized dynamics in the network follows that of (Ahn et al., 2011; Ahn and Rubchinsky, 2017) and is similar to the analysis of the temporal patterns of neural synchrony in the experimental studies mentioned in the Introduction. We will briefly describe this analysis here.

The phase, $\varphi(t)$, of a neuron is defined as

$$\varphi(t) = \tan^{-1}\left(\frac{v(t) - \hat{v}}{w(t) - \hat{w}}\right)$$

where $(\hat{w}, \hat{v})$ is a point selected inside the neuron's limit cycle in the $(w, v)$ – plane. The synchronization strength is computed as

$$\gamma = \left| \frac{1}{N} \sum_{j=1}^{N} exp\left( i\Delta\varphi(t_j) \right) \right|$$

where $\Delta\varphi(t_j) = \varphi_1(t_j) - \varphi_2(t_j)$ is the difference of the phases of neurons 1 and 2 at time $t_j$. $N$ is the number of data points. The value of $\gamma$ ranges from 0 to 1, which represent a complete lack of synchrony and perfect phase synchrony, respectively.

If there is some degree of phase locking present, then there is a synchronized state, i.e. a preferred value of the phase difference $\Delta\varphi$. For each cycle of oscillation one can check if the actual phase difference is close to this preferred value or not. Note that the index $\gamma$ only represents an average value of phase-locking over the interval $[t_1, t_N]$, however to describe the patterning of synchrony one needs to look at the transitions to and from a synchronized state on much shorter timescales. This is done as follows.

When $\varphi_1$ increases past zero, say at time $t_{j,i}$, then $\varphi_2(t_{j,i})$ is recorded. This generates a sequence of numbers $\{\varphi_2(t_{j,i})\}_{i=1}^{M}$. Due to the presence of some synchrony, there is a clustering about some phase value, say $\varphi_0$. This is taken as the preferred phase value, and if $\varphi_2(t_{j,i}) = \varphi_i$, for $1 \leq i \leq M$, differs from it by more than $\frac{\pi}{2}$ then the neurons are desynchronized, otherwise they are synchronized. The choice of $\frac{\pi}{2}$ is not only convenient (it partitions the $(\varphi_i, \varphi_{i+1})$ space into quadrants) but was also used in the experimental studies described in the Introduction.

The length of a desynchronization event is defined as the number of consecutive times the system spends in the desynchronized states. In other words, the length of desynchronization is the length of the time interval the system is away from the synchronized state (as defined above); this length is measured not in the absolute time units, but in the number of cycles of oscillations (in line with the experimental studies mentioned in the Introduction). The lengths of all desynchronization events are recorded and the distribution of durations is reconstructed. The mode of this distribution is used as a characteristic of the temporal patterning of synchronized dynamics. For later reference, a "mode $n$" system means that the mode of all lengths of desynchronization events for that particular system is $n$. Thus a mode 1 system ($n = 1$ case) is the system with synchronized dynamics interrupted by predominantly short desynchronization intervals. The larger $n$ is, the more prominent the tendency for long desynchronizations is. This does not necessarily affect the overall synchrony strength, because it depends not only on the duration of desynchronizations, but also on their number. The mode is used to characterize the durations because experimental studies used the mode for this purpose.

An illustration of different desynchronization durations and dynamics with different modes of desynchronizations is provided in Figure 1. Voltages and distributions of desynchronization durations for mode 1 dynamics are in the left column, the ones for mode 2 dynamics are in the right column. The synchronization is not perfect and synchronized dynamics (phase difference is close to the preferred one) are interspersed with desynchronized intervals. Note that the preferred phase difference is not necessarily zero so that the zero lag state is not necessarily a synchronized state.

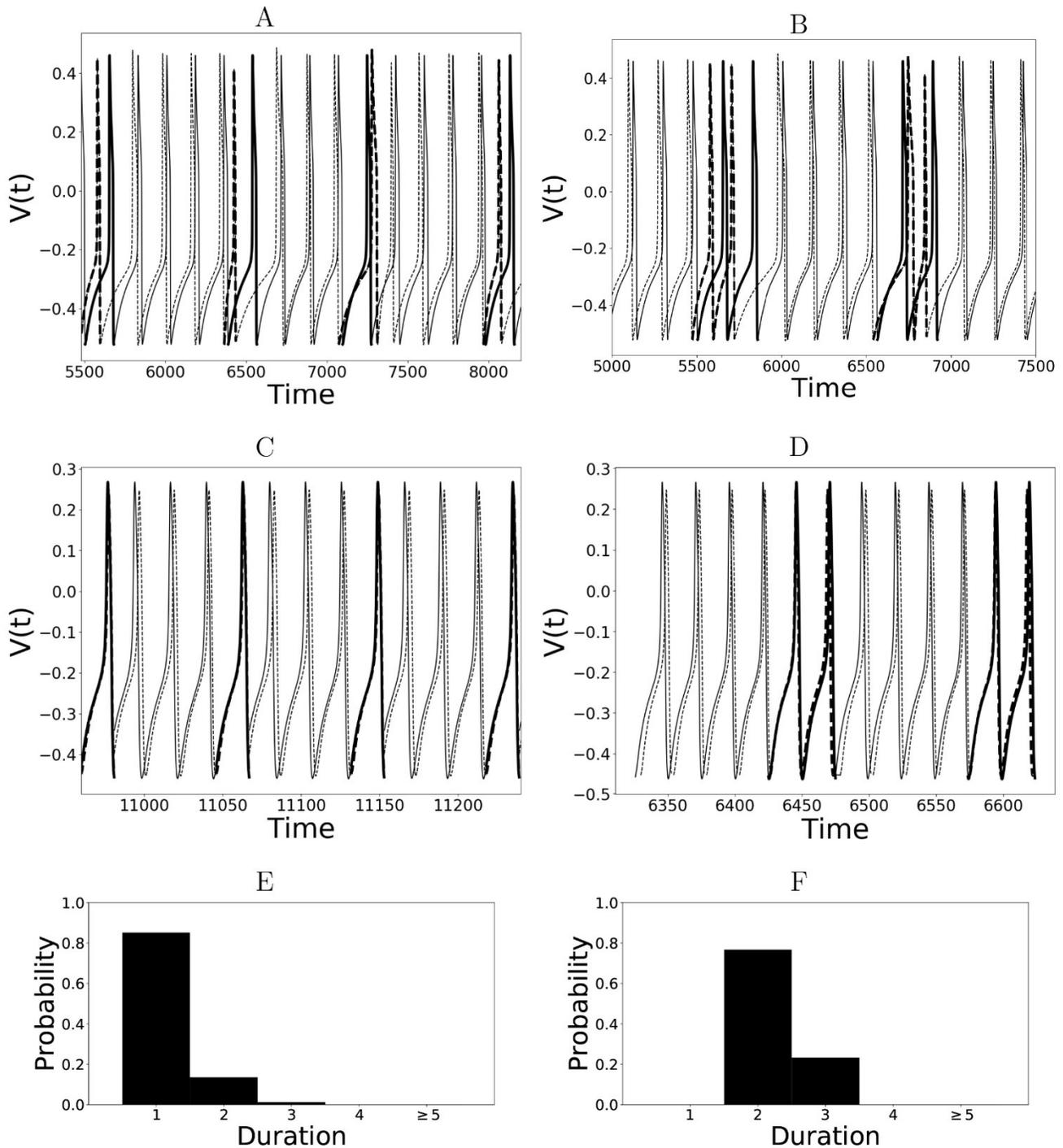

**Figure 1.** Illustration of dynamics with different desynchronization durations (mode 1 and mode 2 dynamics). A–D depict voltage traces of two partially synchronized neurons (solid and dashed lines). When the neurons exhibit the preferred time difference the voltage traces are thin lines, indicating proximity to a synchronized state. However, when the phase difference is not close to the preferred one, the lines are thick to indicate the desynchronizations (as defined above). A and C illustrate short desynchronizations (lasting one cycle of oscillations), B and D show longer desynchronizations (lasting two cycles of oscillations). A and B are artificially generated examples, while C and D present examples generated by the network considered in the section below. In a longer time-series, the desynchronizations of different durations may coexist, however, usually one duration will prevail. The distributions showing relative frequency of different desynchronizations for the dynamics with predominantly short desynchronizations (like A and C) and with longer desynchronizations (like B and D) are presented in E and F respectively. The mode of the distribution in E is 1, thus this is mode 1 dynamics; the mode of the distribution in F is 2, thus this is mode 2 dynamics.

Finally, we would like to reiterate that in this approach the time is measured in terms of cycles of oscillations of the neural activity, not in absolute time units. This allows one to compare the properties of variability of synchrony of brain rhythms with different frequencies.

The phase-locking strength index $\gamma$ was observed to be usually about 0.2-0.3 in this study (even after STDP adjustments). These are moderate values, comparable with experimental results (in particular with the results reported in the studies references in the Introduction). With this moderate synchrony strength, synchronization effects are hard to see by the naked eye, however, the quantitative time-series analysis techniques are able to quantify the synchronized dynamics and its properties including the temporal patterning of weakly synchronous dynamics.

## 3    Results

Building on (Ahn and Rubchinsky, 2017), we used a simple network consisting of two neurons connected via excitatory synapses (see Figure 2); however the synapses are now plastic. The two neurons have a slightly different firing rate, i.e. their respective $\varepsilon$ values differ slightly (see the list of parameter values in Methods). The initial value of the maximal synaptic conductance is $g_{syn} = 0.005$, so that the coupling is weak. This heterogeneity and weak synaptic coupling ensure that the synchrony between the two neurons is relatively weak.

The dynamics of the non-plastic variant of this system was studied in (Ahn and Rubchinsky, 2017). Based on that study, we vary values of three parameters of the potassium current in such a way as to change the dynamics of the non-plastic network from exhibiting predominantly short desynchronizations (i.e. those observed in experiments) to one with a large mode of desynchronization durations. These parameters are $\varepsilon$ (the reciprocal of the peak value of the activation time-constant $\tau(v)$), $\beta$ (which characterizes the widths of the activation time-constant $\tau(v)$ and the steady-state function $w_\infty(v)$) and $v_{w1}$ (a horizontal translation in $w_\infty(v)$ and $\tau(v)$ which changes their values over the specific voltage range). Changes in all these parameters effectively change the activation time-constant $\tau(v)$ to either large or small, which delays or accelerates the activation of potassium current, respectively. Consequently, the lengths of the desynchronization events shift to predominantly short or long. Next, we explore how the introduction of plasticity affects the durations of desynchronization events. Hence our parameter space is two-dimensional for each case considered, and consists only of the plasticity parameters $A$ and $k$.

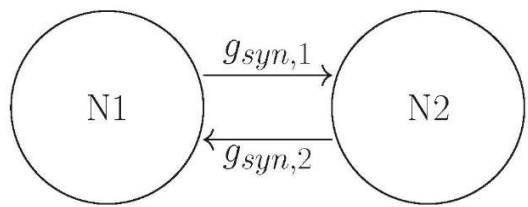

**Figure 2.** The schematics of the network: two neurons coupled with mutually excitatory synapses.

In most of the simulations the synaptic weights do not reach a steady state, but rather exhibit fairly stationary variations, as illustrated in Figure 3.

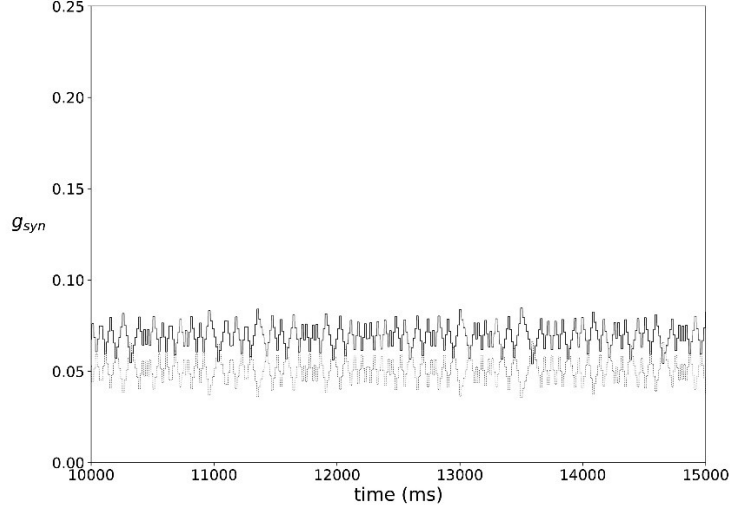

**Figure 3.** An example of typical temporal evolution of synaptic weights in a network with plasticity ($\varepsilon = 0.15$, $A = 0.009$, $k = 0.3$).

## 3.1 Variation of ε

Let us mention here that $\varepsilon \propto \frac{1}{\tau}$ and the maximum value of $\tau(v)$ is $\frac{1}{\varepsilon}$. Hence as $\varepsilon$ is increased, the value of $\tau(v)$ is decreased across its entire domain as it is a unimodal function. This in turn accelerates the activation of potassium current because $\frac{dw}{dt} \propto \frac{1}{\tau(v)}$. From (Ahn and Rubchinsky, 2017) we know that smaller values of $\varepsilon$ promote shorter desynchronization events.

For $\varepsilon = 0.05$, the non-plastic system is mode 1. This means the synchronized dynamics has the following property. As the system is exhibiting partially synchronized dynamics, it will be either close in the synchronized state or away from synchronized state, the latter is termed desynchronization. The desynchronized interval length (measured in the number of cycles of oscillations) varies in time. We obtain the distribution of the desynchronization durations from numerical simulation and find the mode of this distribution. If this mode equals one cycle of oscillation, then the system is mode 1 (see Methods for a more detailed explanation). Mode 1 means the desynchronizations are predominantly short.

Now the non-plastic system is changed to include STDP. The changes in the temporal patterning of synchronization dynamics are illustrated in Figure 4. Figure 4A is a diagram of the mode of the desynchronization durations in the space of plasticity parameters, $A$ and $k$. The plasticity effects are negligible across the top (very large $k$ implies a quick decay of the change in synaptic strength), and especially in the upper left corner (large $k$ and a small amplitude $A$). In these areas the values of the plasticity parameters are such that the magnitude of the update, $\Delta g_{syn}$, is negligible (the average update is usually in the interval $[0.0, 10^{-5}]$, on the larger end this corresponds to about $0.2\%$ of the initial value of $g_{syn}$). Hence, the plastic system continues to be mode 1 in these areas.

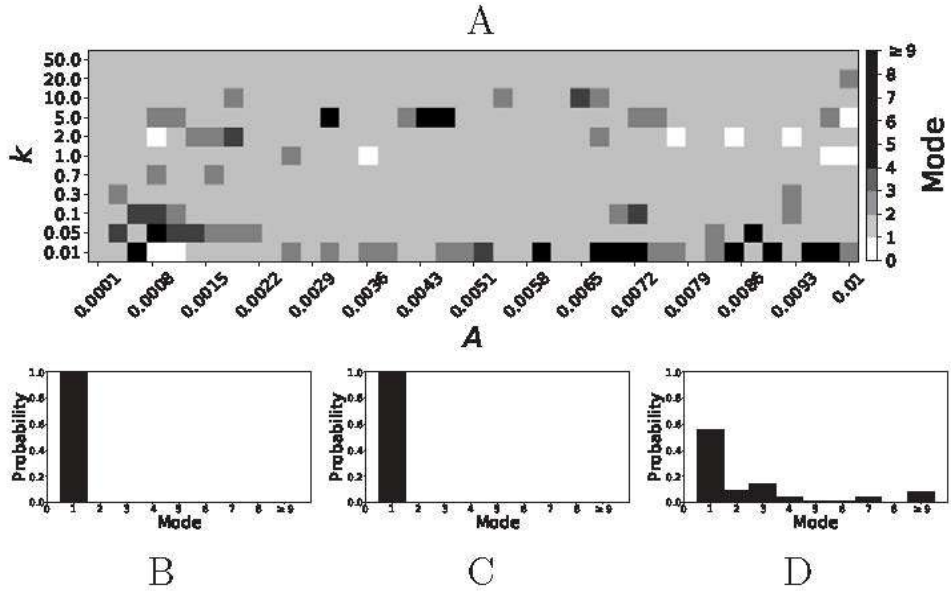

**Figure 4.** A system exhibiting mode 1 dynamics in the non-plastic case is subjected to plasticity ($\varepsilon = 0.05$). A: Mode is colored via gray scale, see legend on the right of the diagram. The amplitude of the synaptic update, $A$, is varied along the horizontal axis. The reciprocal of the time-scale of the synaptic update, $k$, is varied along the vertical axis. B, C and D show the changes in the histogram of desynchronization durations as plasticity becomes stronger. B: The system without plasticity. C: The system with very weak plasticity: $A = 0.0047, k = 20.0$. D: The system with moderate plasticity: $A = 0.0047, k = 0.05$.

The rest of the parameter space, in particular the central region, displays a high proportion of mode 1 dynamics as well. In these areas plasticity is not negligible, as the synaptic strength can vary to a substantial degree. However, even in the presence of STDP, mode 1 dynamics persist. For the diagram in Figure 4A, about 85% of the parameter space points correspond to mode 1 systems.

To illustrate the effect of plasticity on a distribution of desynchronization durations, refer to Figure 4B, 4C and 4D. Plasticity effects increase from left to right. The distribution of durations changes: at a weak level of plasticity the durations are exclusively length one, while at a stronger level of plasticity some longer durations are observed. Yet the preponderance of length one desynchronization durations is preserved.

Now let us look at the effect of plasticity on the dynamics in systems with a mode larger than one. We consider $\varepsilon = 0.15$. The non-plastic system is mode 2 (the synchronization index $\gamma$ is virtually unchanged from that of $\varepsilon = 0.05$, although the frequency of oscillations increases by several times,

Ahn and Rubchinsky, 2017). Mode 2 means the desynchronizations tend to be longer than those of the mode 1 case.

Figure 5 shows the effect of STDP on the system that is mode 2 in the non-plastic case. As explained earlier, the plasticity effects are negligible across the top of Figure 5A, and especially in the upper left corner. We note that this region of the parameter space exhibits mode 2 dynamics (as expected). However, throughout the entire parameter space it is seen that a majority of parameter values correspond to mode 1 systems (the large central region in Figure 5A). Overall, about 20% of the parameter space points stay mode 2, while over 65% exhibit mode 1 dynamics (and less than 15% correspond to larger than mode 2 systems).

To illustrate the effect of plasticity on a distribution of desynchronization durations, refer to Figure 5B, 5C and 5D. Plasticity effects increase from left to right. Here we see that the introduction of weak plasticity can be sufficient to shift the system from mode 2 to mode 1 (Figure 5C). This means desynchronizations tend to become shorter in the plastic case. At stronger levels of plasticity (Figure 5D), the distribution widens, however the vast majority of desynchronization events remain length one.

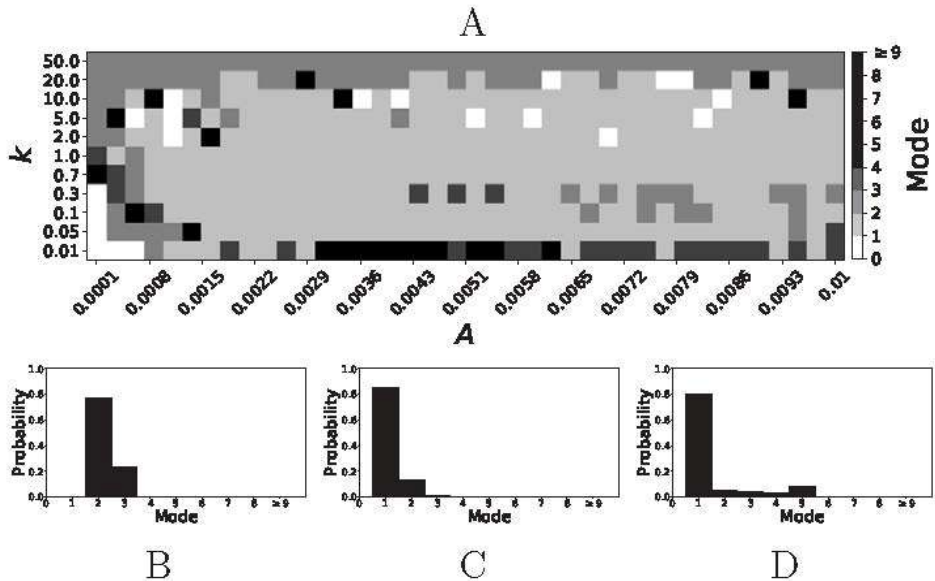

**Figure 5.** A system exhibiting mode 2 dynamics in the non-plastic case is subjected to plasticity ($\varepsilon = 0.15$). A: Mode is colored via gray scale, see legend on the right of the diagram. The amplitude of the synaptic update, $A$, is varied along the horizontal axis. The reciprocal of the time-scale of the synaptic update, $k$, is varied along the vertical axis. B, C and D show the changes in the histogram of desynchronization durations as plasticity becomes stronger. B: The system without plasticity. C: The system with very weak plasticity: $A = 0.0047, k = 20.0$. D: The system with moderate plasticity: $A = 0.0047, k = 0.7$.

Overall, we have seen that mode 1 dynamics are generally preserved when STDP is introduced to a non-plastic mode 1 system. When STDP is introduced to a non-plastic mode 2 system, the dynamics largely shifts from mode 2 to mode 1. The same was found with other non-plastic systems exhibiting higher modes: the introduction of STDP generally shifts the mode of the system down to one. Finally, we would like to note that there are several points in the parameter space (see Figure 4A and Figure 5A) that have very large modes. For example, in Figure 4A when $A = 0.0006, k = 0.01$, the resulting system is mode 38 (i.e. most common desynchronizations are very long). Generally, these cases have a wide distribution of desynchronization durations. Therefore, while these systems have a large mode, the mode does not present a strong tendency in the distribution. Nevertheless, these situations are relatively rarely found.

### 3.2 Variation of β

The parameter $\beta$ changes the widths of the voltage-dependent time-constant of activation $\tau(v)$ and the width of the steady-state activation function $w_\infty(v)$ for potassium current. In particular, as $\beta$ is decreased, the slope at the half-height of $w_\infty(v)$ is increased, and this decreases the width of the step ($w_\infty(v)$ is a sigmoidal function). Similarly, for $\tau(v)$, a decrease in $\beta$ decreases the width of the function around the peak. This causes an advancement in the activation of the potassium current.

A larger value of $\beta$ promotes shorter desynchronization durations (Ahn and Rubchinsky, 2017). For $\beta = 0.124$, the non-plastic system is mode 1. The effect of STDP on this system is presented in Figure 6. Across the top and in the upper left corner of Figure 6A we see that virtually every point corresponds to a mode 1 system, as expected. Indeed, a substantial portion of the entire parameter space displays mode 1 dynamics; about 80% of the parameter space studied.

To illustrate the effect of plasticity on a distribution of desynchronization durations, refer to Figure 6B, 6C and 6D. Plasticity effects increase from left to right. The introduction of plasticity has a minimal effect on the distribution; there is very little change visibly. Indeed, the proportion of desynchronization durations of length one increases with plasticity.

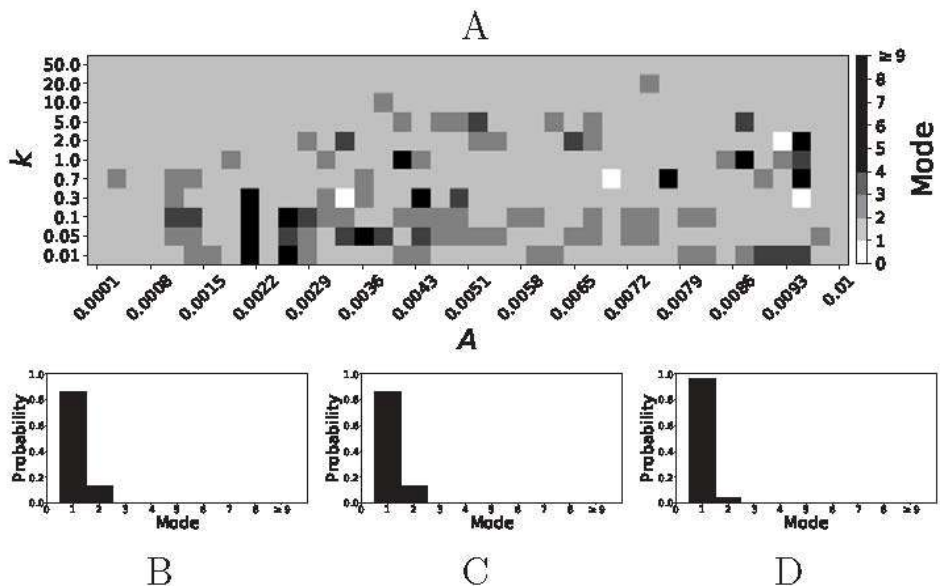

**Figure 6.** A system exhibiting mode 1 dynamics in the non-plastic case is subjected to plasticity ($\beta = 0.124$). A: Mode is colored via gray scale, see legend on the right of the diagram. The amplitude of the synaptic update, $A$, is varied along the horizontal axis. The reciprocal of the time-scale of the synaptic update, $k$, is varied along the vertical axis. B, C and D show the changes in the histogram of desynchronization durations as plasticity becomes stronger. B: The system without plasticity. C: The system with very weak plasticity: $A = 0.0052, k = 20.0$. D: The system with moderate plasticity: $A = 0.0052, k = 0.7$.

Decreasing $\beta$ increases the mode of a system. If $\beta = 0.091$, the non-plastic system is mode 2. With the introduction of very weak plasticity (across the top and the upper left corner of Figure 7A) we see that the dynamics are relatively unchanged, i.e. the mode of most systems remains two. However, if plasticity is not very weak, the dynamics shift to mode 1 in a significant portion of the parameter space. The effect is not as substantial as in the previous section, but about 35% of parameter space becomes mode 1 (about 45% remains mode 2, i.e. the mode is unchanged).

To illustrate the effect of plasticity on a distribution of desynchronization durations, refer to Figure 7B, 7C and 7D. Plasticity effects increase from left to right. We see that the vast majority of desynchronization durations become length one as plasticity becomes stronger.

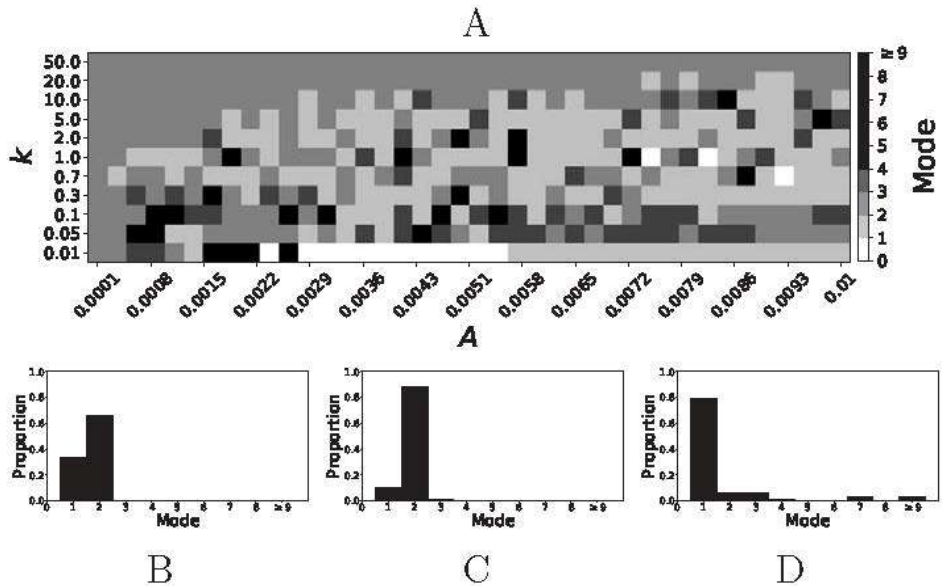

**Figure 7.** A system exhibiting mode 2 dynamics in the non-plastic case is subjected to plasticity ($\beta$=0.091). A: Mode is colored via gray scale, see legend on the right of the diagram. The amplitude of the synaptic update, $A$, is varied along the horizontal axis. The reciprocal of the time-scale of the synaptic update, $k$, is varied along the vertical axis. B, C and D show the changes in the histogram of desynchronization durations as plasticity becomes stronger. B: The system without plasticity. C: The system with very weak plasticity: $A = 0.0047, k = 20.0$. D: The system with moderate plasticity: $A = 0.0047, k = 0.7$.

### 3.3 Variation of $v_{w1}$

The parameter $v_{w1}$ affects a horizontal translation in $w_\infty(v)$ and $\tau(v)$. Increasing $v_{w1}$ shifts both curves to the right, i.e. towards higher voltages; this results in a potassium current that activates faster.

Smaller values of $v_{w1}$ result in short desynchronization durations (Ahn and Rubchinsky, 2017). For $v_{w1} = 0.102$, the non-plastic system is mode 1. The effect of STDP on this system is presented in Figure 8. We see that mode 1 dynamics is observed not only for the weak plasticity region (top and upper left corner of Figure 8A), but for most of the parameter space (about 85% of the parameter space studied).

To illustrate the effect of plasticity on a distribution of desynchronization durations, refer to Figure 8B, 8C and 8D. Plasticity effects increase from left to right. We see that as plasticity increases to a higher level, the prevalence of mode 1 is unchanged.

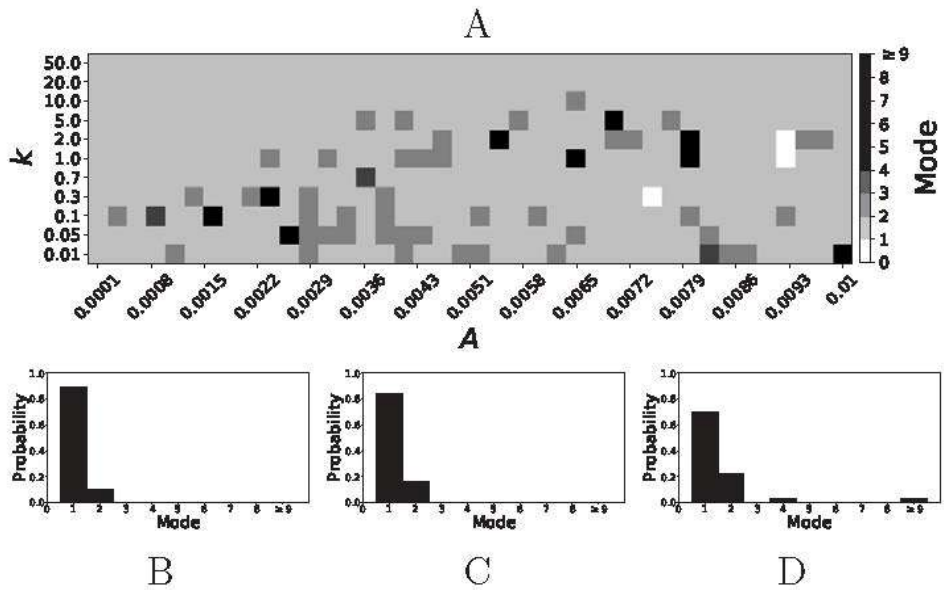

**Figure 8.** A system exhibiting mode 1 dynamics in the non-plastic case is subjected to plasticity ($v_{w1} = 0.102$). A: Mode is colored via gray scale, see legend on the right of the diagram. The amplitude of the synaptic update, $A$, is varied along the horizontal axis. The reciprocal of the time-scale of the synaptic update, $k$, is varied along the vertical axis. B, C and D show the changes in the histogram of desynchronization durations as plasticity becomes stronger. B: The system without plasticity. C: The system with very weak plasticity: $A = 0.0047, k = 20.0$. D: The system with moderate plasticity: $A = 0.0047, k = 0.7$.

Varying $v_{w1}$ to larger values leads to shorter desynchronization durations becoming less prevalent. For $v_{w1} = 0.161$, the non-plastic system is mode 2. The effect of STDP is presented in Figure 9. When plasticity is added we see that the dynamics are similar to the non-plastic case when plasticity is weak enough (top and upper left corner of Figure 9A). However, when the plasticity effects are moderate, the system exhibits mode 1 dynamics frequently (central region of Figure 9A). For the domain of parameter space studied, the majority of points (about 45%) correspond to mode 1 systems, the rest are either mode 2 (about 40%) or higher.

To illustrate the effect of plasticity on a distribution of desynchronization durations, refer to Figure 9B, 9C and 9D. Plasticity effects increase from left to right. We see that the mode of the system shifts down from two to one as plasticity becomes stronger.

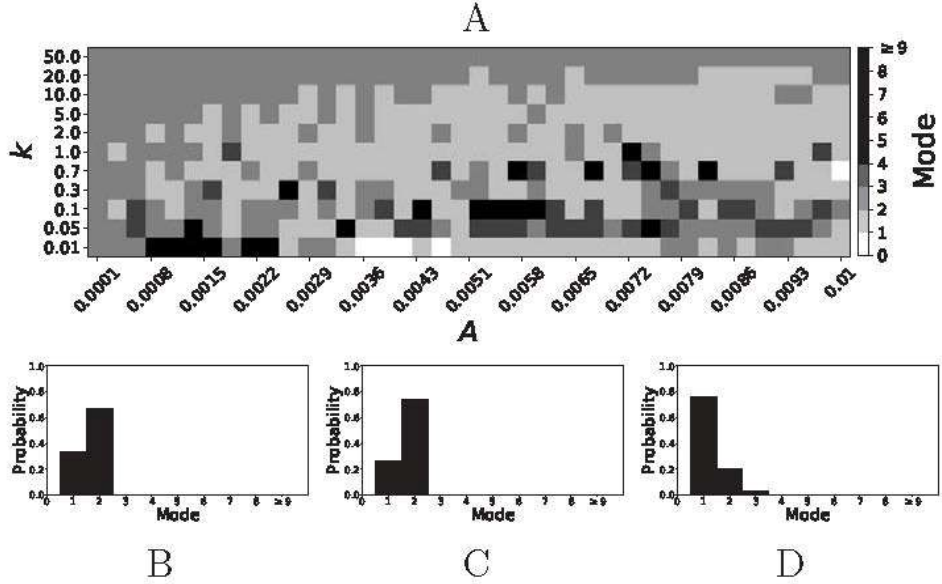

**Figure 9.** A system exhibiting mode 2 dynamics in the non-plastic case is subjected to plasticity ($v_{w1} = 0.161$). A: Mode is colored via gray scale, see legend on the right of the diagram. The amplitude of the synaptic update, $A$, is varied along the horizontal axis. The reciprocal of the time-scale of the synaptic update, $k$, is varied along the vertical axis. B, C and D show the changes in the histogram of desynchronization durations as plasticity becomes stronger. B: The system without plasticity. C: The system with very weak plasticity: $A = 0.0047, k = 20.0$. D: The system with moderate plasticity: $A = 0.0054, k = 1.0$.

### 3.4   Variation of $\beta_w$ and $\beta_\tau$

Varying either $\varepsilon$, $\beta$ or $v_{w1}$ may affect not only the durations of the desynchronizations, but also synchronization strength and the frequency of activity in the system. To control desynchronization durations while keeping both spiking frequency and synchronization strength near constant in a non-plastic system, one can consider the parameter $\beta$ and separate it into two independent parameters, $\beta_\tau$ and $\beta_w$. As a result, the lengths of desynchronization events are almost independent of the frequency and synchrony strength (Ahn and Rubchinsky, 2017).

Smaller $\beta_w$ and larger $\beta_\tau$ result in shorter desynchronization durations (Ahn and Rubchinsky, 2017). For $\beta_w = 0.098, \beta_\tau = 0.079$, the non-plastic system is mode 1. Figure 10 illustrates the impact of STDP on this system. Mode 1 dynamics is observed not only for the weak plasticity region (top and upper left corner of Figure 10A), but for the majority of the parameter space (about 60% of the parameter space studied).

To illustrate the effect of plasticity on a distribution of desynchronization durations, refer to Figure 10B, 10C and 10D. Plasticity effects increase from left to right. We see that as plasticity progresses to a moderate level, the proportion of short desynchronizations stays largely unchanged. In particular, the system is still mode 1.

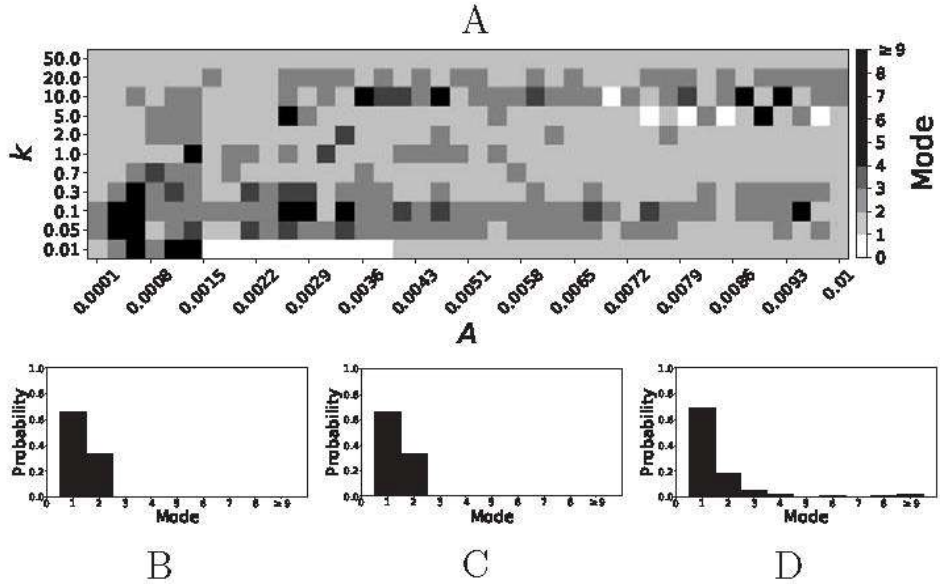

**Figure 10.** A system exhibiting mode 1 dynamics in the non-plastic case is subjected to plasticity ($\beta_w = 0.098, \beta_\tau = 0.079$). A: Mode is colored via gray scale, see legend on the right of the diagram. The amplitude of the synaptic update, $A$, is varied along the horizontal axis. The reciprocal of the time-scale of the synaptic update, $k$, is varied along the vertical axis. B, C and D show the changes in the histogram of desynchronization durations as plasticity becomes stronger. B: The system without plasticity. C: The system with very weak plasticity: $A = 0.0049, k = 50.0$. D: The system with moderate plasticity: $A = 0.0052, k = 0.7$.

If $\beta_w = 0.115, \beta_\tau = 0.071$, the non-plastic system is mode 2. Figure 11 illustrates the impact of STDP on this system. With the addition of plasticity, we see that the system is largely mode 2 if the plasticity is weak (top and upper left corner of Figure 11A). However, stronger plasticity shifts the dynamics to mode 1 for a substantial portion of the parameter space (about 55% of points considered).

To illustrate the effect of plasticity on a distribution of desynchronization durations, refer to Figure 11B, 11C and 11D. Plasticity effects increase from left to right. We see that the distribution is largely unchanged for very weak plasticity, but as plasticity increases, the system becomes mode 1.

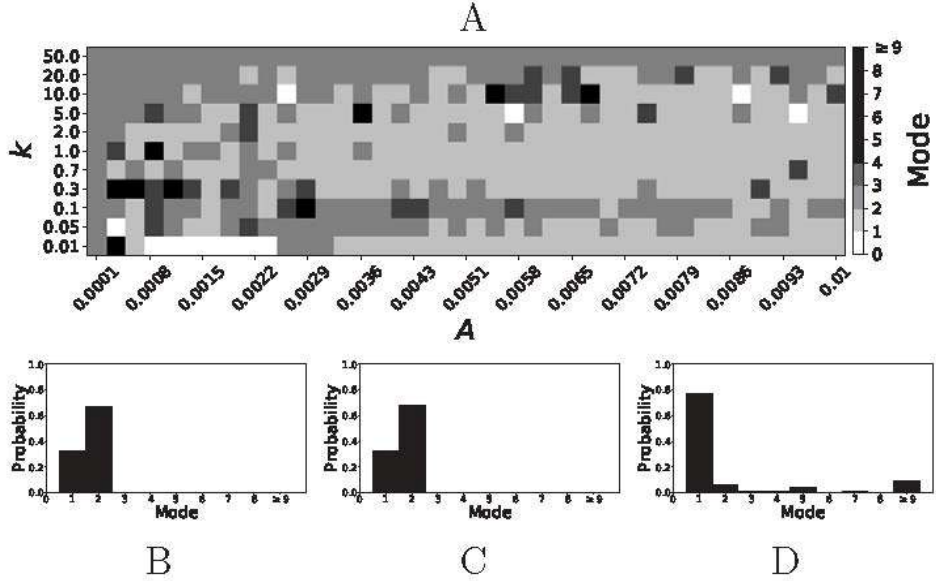

**Figure 11.** A system exhibiting mode 2 dynamics in the non-plastic case is subjected to plasticity ($\beta_w = 0.115, \beta_\tau = 0.071$). A: Mode is colored via gray scale, see legend on the right of the diagram. The amplitude of the synaptic update, $A$, is varied along the horizontal axis. The reciprocal of the time-scale of the synaptic update, $k$, is varied along the vertical axis. B, C and D show the changes in the histogram of desynchronization durations as plasticity becomes stronger. B: The system without plasticity. C: The system with very weak plasticity: $A = 0.0049, k = 50.0$. D: The system with moderate plasticity: $A = 0.0054, k = 0.7$.

## 4    Discussion

This study considered intermittent synchronous dynamics in a small network of simple conductance-based model neurons. While strong synaptic strength in general can promote synchronization between neurons, moderate values of synaptic coupling lead to dynamics with relatively weak synchronization, and where the episodes of synchronization are interspersed with episodes of desynchronized dynamics. Intermittent synchronization in the presence of moderate (and fixed in time) coupling is quite typical for coupled oscillatory systems (Pikovsky et al., 2001). In other words, temporal variability of correlations is observed due to the relative weakness of a fixed coupling strength. The temporal signatures of this variability have been previously modeled in (Ahn and Rubchinsky, 2017) and were in good agreement with the analysis of the temporal variability observed in experimental data (see Introduction and references therein).

However, many actual synapses are plastic and thus the synaptic coupling between neurons experiences temporal variations. This variation may contribute to the temporal variability of intermittent synchrony as well. This study considered how one common type of neural plasticity – spike-timing dependent plasticity – might affect this temporal variability. Experimental data ubiquitously points to the prevalence of short desynchronization dynamics in neural synchrony. This kind of dynamics is naturally generated in synaptically coupled conductance-based model neurons. We showed here that the introduction of STDP under quite general conditions preserves this realistic fine temporal structure of intermittent neural synchrony. Moreover, when the non-plastic system parameters are selected in such a way as to predominantly express longer desynchronizations, STDP changes the intermittently synchronous dynamics back to one with short desynchronizations. This was observed while varying several different parameters, so that STDP may reverse dynamics from long to short desynchronizations regardless of how the desynchronizations were obtained in the non-plastic system.

The overall dependence of the dynamics on the characteristics of plasticity is quite complicated. Numerical simulations indicate that some plasticity parameter values may promote very unrealistic synchronized dynamics. However, under the conditions considered, the short desynchronization dynamics were obtained in large regions of the parameter space. This was regardless of whether the corresponding non-plastic system was mode 1, or had a higher mode.

The results of these numerical simulations suggest that STDP may be one of the contributing factors behind experimentally observed short desynchronization dynamics. Moreover, STDP and cellular mechanisms proposed in (Ahn and Rubchinsky, 2017) may act cooperatively in promoting short desynchronizations.

The results discussed here were obtained in the framework of relatively simple modeling. The actual neuronal synchrony is, of course, a much more complicated phenomenon than the model considered here, and there were multiple factors not included in the model. For example, inhibitory synapses (e.g. see Nowotny et al., 2008). The experimental observations of short desynchronizations were mostly done with LFP and EEG signals, and the simple network considered here is too simple to adequately model these signals. However, the similarity between experimentally observed intermittent neural synchrony and the temporal patterning of synchrony observed in our study with a relatively simple model with STDP may speak to the very general nature of this phenomenon.

The variability of the dynamics on short time-scales may be a functionally beneficial phenomenon. Short desynchronization dynamics (which is essentially a high degree of variability of synchrony on very short time-scales) have been conjectured to be conducive for quick and efficient formation and break-up of neural assemblies (Ahn and Rubchinsky, 2013, 2017). As was noted in these studies, the ease of formation and disappearance of synchronized states at rest may suggest that a transient synchronized assembly may be easily formed whenever needed to facilitate a particular function. The results of this study suggest that the temporal variability of synaptic strength due to STDP may potentially further facilitate this phenomenon.

## 5    Conflict of Interest

The authors declare that the research was conducted in the absence of any commercial or financial relationships that could be construed as a potential conflict of interest.

## 6    Author Contributions

LR conceived research; LR and JZ designed research; JZ performed numerical simulations, JZ and LR analyzed and interpreted the results; JZ and LR wrote the manuscript.

## 7    Funding

Supported by NSF DMS 1813819.